\documentclass[10pt]{article}

\usepackage{
graphicx,amsmath,amssymb,bm}

\title{Uniqueness of Ground state  \\in the  Edwards-Anderson  Spin Glass Model}
\author{C. Itoi\\
Department of Physics, GS $\&$ CST, Nihon University, \\
Kandasurugadai, Chiyoda,  Tokyo 101-8308, Japan}

\begin{document}
\maketitle 
\begin{abstract}It is proven rigorously  that the ground state in  the Edwards-Anderson spin glass model is unique
  in any dimension  for almost all continuous random exchange interactions 
  under a condition that
  a single spin  breaks the global  ${\mathbb Z}_2$ symmetry.
This theorem implies that replica symmetry breaking does not occur at zero temperature.  
The site- and  bond-overlap are concentrated at their maximal values. 
 Behaviors of short range spin glass models  turn out to be 
  much different  
 from those of  mean field spin glass models near zero temperature.  Errata have been attached at the final page.
\end{abstract}

\maketitle



\paragraph{Introduction}
Recent progress in theoretical physics of spin glass models is remarkable. The replica symmetry breaking phenomena 
 in mean field spin glass models have been understood deeply, since Talagrand proved  the Parisi formula \cite{Pr} for the free energy of the
Sherrington-Kirkpatrick (SK) model \cite{SK} in a mathematically rigorous manner \cite{T,T2}. The replica symmetry breaking (RSB)  phase includes the 
spin glass phase in the SK model. 
RSB is a nontrivial phenomenon which is expected to occur generally in many disordered systems
due to their infinitely  many low lying states with infinitesimal energy gaps including ground states. Recently, this fact is proven rigorously \cite{ACZ}.  
Theoretical physicists and mathematicians have been  seeking this phenomenon also in more realistic  short range spin glass models, such as the Edwards-Anderson (EA) model \cite{EA}. They have been devoting great deal of attention to whether or not, RSB occurs in such models as well as in mean field spin glass models. 
The RSB picture proposed by Parisi claims that there are many pure states  also in the short range spin glass models
as in mean the SK model, and RSB appears \cite{P2}. 
On the other hand, Fisher and Huse  have argued the absence of many pure states in the short range spin glass models
on the basis of the droplet picture  \cite{FH}.
There have been many numerical simulations in the three dimensional EA model. 
Some of them suggest the RSB picture for the three dimensional EA model \cite{MPRRZ,KPY, BCFGM,WWL}, while  some others  deny the RSB picture \cite{YK, YKM}
. 
 It is believed that the critical phenomena in 
short-range interacting spin models behave like those in  mean field spin model  in  any dimensions 
higher than the upper critical dimension. In  this sense, the RSB should be observed 
also in  short-range interacting spin models in some higher dimensions.  
Although this controversial question in theoretical physics has been  argued for four decades after the discovery of the
Parisi formula \cite{Pr} for the SK  model, there has never been any clear answer except in the limited studies. 
There are a few rigorous results for  RSB in low temperature region of short range disordered spin models.
Nishimori and Sherrington have shown that the RSB does not
occur on the Nishimori line which is located out of the spin glass phase in the EA model \cite{NS1,N}.  
Arguin and Damron  have proven rigorously  that  the number of ground states in the EA model is either 2 or $\infty$ with probability one on the half-plane \cite{AD}.
Within this result, it is unclear whether or not RSB occurs. 
Recently, Chatterjee has proven  a remarkable theorem that the random field Ising model has no extended RSB phase in any dimension \cite{C2}.  It has been shown that the distribution of the overlap is concentrated at a single value given by its expectation value almost everywhere in the coupling constant space, unlike mean field spin glass models.    
This theorem   is proven by utilizing three key tools:  the Fortuin-Kasteleyn-Ginibre (FKG) inequality \cite{FKG},
the Ghirlanda-Guerra identities\cite{GG,AC} and the Chatterjee inequalities \cite{C2}. 
While the Ghirlanda-Guerra identities are well-known to hold  universally in wide class of spin systems
with Gaussian random interactions,  
the FKG inequality is valid only in the random field Ising model  with positive definite exchange interactions
\cite{FKG}. This result has established rigorously the already argued
claim that the random field Ising model has no spin glass phase \cite{K}.

In the present paper, it is  proven that  the 
EA model has  a unique ground state  in any dimension 
 for almost all continuous random exchange interactions under a condition that a single spin
  breaks the global ${\mathbb Z}_2$ symmetry. 
This implies that RSB does not occur 
in the EA model with continuous random exchange interactions at zero temperature in any finite dimensions.
A similar method to a recently developed fundamental method 
\cite{I} enables us to prove 
 this important property of short-range  disordered Ising systems,
 without using the FKG and Chatterjee inequalities and the Ghirlanda-Guerra identities.
 The proof uses  simple and elementary methods.
To argue the absence of RSB  also near zero temperature,  we consider
the property of energy gap of a spin configuration above the unique ground state.
Behaviors of short range spin glass models  turn out to be much different  
 from those of  mean field spin glass models. 
The long-standing unsolved problem is solved 
in the present paper. \\

\paragraph{Definitions and main theorem} 
Consider $d$-dimensional hyper cubic lattice  $\Lambda_L= {\mathbb Z}^d \cap [-L,L]^d$
with a positive  integer $L$. 
  Define a set of nearest neighbor bonds by
 $B_L = \{ \{i, j\}|  i,j \in \Lambda _L, |i-j|=1\}$. Note $|B_L|=|\Lambda_L| d$.
 Let  $\Sigma_L:=\{-1,1\}^{\Lambda_L}$ be a set of  spin configurations $\sigma : \Lambda_L \to \{-1,1\}$.
 Let $\bm J=(J_{ij})_{\{i,j\} \in B_L}$ be a sequence of 
 independent and identically distributed (i.i.d) continuous random variables, 
whose expectation value and variance are given by  
\begin{eqnarray}
&&{\mathbb E} J_{ij}=J_0, \ \   {\mathbb E} (J_{ij}-J_0)^2=J^2, 
\end{eqnarray} 
for $J>0$ and  $J_0\in {\mathbb R}$.
The Hamiltonian of this model
\begin{equation}
H_{L}(\sigma, \bm J)= - \sum_{\{i,j\} \in B_L} J_{ij}\sigma_i \sigma_j
,
\end{equation}
 is  a function of  spin configuration  $\sigma \in  \Sigma_L$
  and a random sequence  $\bm J.$
  For any $\beta >0 $,  the  partition function as a function of   $(\beta, h,  \bm J)$ is defined by
\begin{equation}
Z_L(\beta, \bm J) = \sum_{\sigma \in \Sigma_L} e^{ - \beta H_L(\sigma, \bm J)},
\end{equation}
with the free boundary condition.
The average of an arbitrary function $f: \Sigma_ L\rightarrow {\mathbb R}
$ 
of the spin configuration in the Gibbs state is given by
$$
\langle f(\sigma) \rangle = \frac{1}{Z_L(\beta, \bm J)}\sum_{\sigma \in \Sigma_L} f(\sigma)  e^{  -\beta H_L(\sigma,\bm J)}.
$$ 
The free energy density as a function of   $(\beta, \bm J)$
is defined by
\begin{equation}
\varphi_L(\beta, \bm J) :=- \frac{1}{|\Lambda_L|\beta} \log Z_L(\beta, \bm J). \\ 
\end{equation}
Uniform convergence of the expectation value of  the free energy density
$$\displaystyle
\lim_{L\to\infty}{\mathbb E}  \varphi_L(\beta, \bm J),
$$ 
can be  proven, where ${\mathbb E}$ denotes expectation over the random variables 
$\bm J$. The self-averaging property of $\varphi_L(\beta, \bm J)$ has been  proven \cite{I}.

 This Hamiltonian is invariant under the action of ${\mathbb Z}_2$ on the spin configuration
 $\sigma\mapsto -\sigma$.  Note that the expectation $\langle \sigma_i\rangle$ of spin
at each site $i$ vanishes  in the ${\mathbb Z}_2$ symmetric Gibbs state. 
To study the spontaneous  symmetry breaking of  the global ${\mathbb Z}_2$ symmetry,
assume  a  condition at  the origin $\bm 0:=(0,0, \cdots, 0)$
\begin{equation}\sigma_{\bm 0}=1,
\label{BC}
\end{equation}  
to remove the two-fold degeneracy. 
The phases are classified  by   the ferromagnetic order parameter for $M<L$
$$\displaystyle m:=\lim_{M\to\infty}\lim_{L\to\infty} \frac{1}{|\Lambda_M|} \sum_{i\in \Lambda_M} \langle 
\sigma_i \rangle,
$$
and the Edwards-Anderson spin glass order parameter 
$$\displaystyle q:= \lim_{M\to\infty}\lim_{L\to\infty}\frac{1}{|\Lambda_M|} \sum_{i\in \Lambda_M} \langle \sigma_i\rangle^2.$$
Note $m ^2 \leq q$. The three phases, 
a  ${\mathbb Z}_2$ broken phase  $ m  \neq 0$, $q\neq 0$, 
another  broken phase $m  = 0,$ $q\neq 0$ and the unique symmetric phase  $ m  =q=0$ define
the ferromagnetic phase, the spin glass phase
and the paramagnetic phase, respectively.

To study replica symmetry, define $n$ replicated spin configurations $(\sigma^1, \cdots, \sigma^n)\in \Sigma_L^n$.
The bond-overlap $R_{k,l}$  and the site-overlap $S_{kl}$ $(1\leq k,l \leq n)$
between $k$-th and $l$-th spin configurations are defined by
\begin{equation}
R_{k,l}=\frac{1}{|B_M|} \sum_{\{i,j\} \in B_M} \sigma_i^k \sigma_j^k  \sigma_i^l \sigma_j ^l, \ \ 
S_{k,l} = \frac{1}{|\Lambda_M|} \sum_{i \in \Lambda_M} \sigma_i^k \sigma_i^l, \label{site}
\end{equation}
The bond-overlap is a function of two replicated spin configurations.
Here, we consider the Hamiltonian as a function of $n$ spin configurations sharing the same random variables $\bm J$
 \begin{equation}
 H(\sigma^1, \cdots, \sigma^n, \bm J
  ):= \sum_{k=1} ^n H_L(\sigma^k, \bm J)
\label{RSBHamil}
\end{equation}
Hamiltonian is invariant under any permutation $s$ among $n$ replicated spin configurations.
$$
 H(\sigma^{s(1)}, \cdots, \sigma^{s(n)}, \bm J
  ) =  H(\sigma^1, \cdots, \sigma^n, \bm J
  )
$$ 
This is called replica symmetry. 
 If we calculate the expectation of the  site-overlap in the replica symmetric Gibbs state,
it is identical to the Edwards-Anderson spin glass order parameter.
$$
\langle S_{k,l} \rangle= \frac{1}{|\Lambda_M|} \sum_{i \in \Lambda_L}\langle  \sigma_i^k \sigma_i^l  \rangle =  \frac{1}{|\Lambda_M|} \sum_{i \in \Lambda_M} \langle \sigma_i \rangle^2 =q,
$$
The distribution of the site-overlap is broadened in a certain low temperature region including spin glass phase in the SK model, 
where the replica symmetric Gibbs state becomes unstable. This phenomenon is  
RSB,  conjectured by Parisi \cite{Pr} for the SK model, and proven by Talagrand \cite{T}. 
The condition (\ref{BC}) enables us to detect the finite variance only due to the RSB without confusion
due to the ${\mathbb Z}_2$ symmetry.
The RSB has been observed  in several mean field models \cite{G2,G,Pn}, while in the short range spin glass model, it has been unclear until now.

There have been several criticism  on the RSB picture
for short range spin models. 
Newman and Stein have claimed that a short range spin glass model should have
a pure Gibbs state, then the RSB picture is unnatural in statistical physics \cite{NS}.  
Uniqueness of the ground state and  non-existence of RSB in the EA model at zero temperature
 are shown by the following theorem,
which confirms the claims of Fisher-Huse and   Newman-Stein. \\

\noindent
{\bf Theorem} {\it   Consider  the  Edwards-Anderson (EA) model in $d$-dimensional hyper cubic lattice $\Lambda_L$
under the  condition  (\ref{BC}).
For  $M < L$,  let   
  $f(\sigma)$ be a real valued  function of a spin configuration
 $\sigma\in \Sigma_M$
 .
For almost all $\bm J$, there 
exists a unique spin configuration $\sigma^+ \in \Sigma_M$,
such that the following limit 
is given by
\begin{eqnarray}
\lim_{\beta \to \infty} \lim_{L \to \infty} \langle f(\sigma ) 
\rangle= f( \sigma^+).
\label{bondexp} 
\end{eqnarray}   
}

This theorem implies the following Corollary that 
 RSB does not occur  in the EA model in  Chatterjee's definition \cite{C2}. 
 Chatterjee defines the absence of RSB in an arbitrary random spin systems, if 
 the variance of the overlap is vanishes
  \begin{eqnarray}
&&\lim_{M \rightarrow \infty }\lim_{\beta \to \infty}   \lim_{L \to \infty}
{\mathbb E} \langle (S_{1,2} -{\mathbb E}\langle S_{1,2} \rangle)^2\rangle=0,\\
&&\lim_{M \rightarrow \infty }\lim_{\beta \to \infty} \lim_{L \to \infty}{\mathbb E} \langle (R_{1,2} -{\mathbb E}\langle R_{1,2} \rangle)^2\rangle=0,
\label{Chatterjee}
\end{eqnarray}
 in the replica symmetric Gibbs state and sample expectation.
 If replica symmetry breaking occurs as a spontaneous symmetry breaking, then the
 variance of the order parameter becomes finite  in the Gibbs state having the corresponding symmetry. 
Chatterjee's definition of  the absence of RSB is contrapositive of the above claim.
The site- and bond-overlap are concentrated at their maximal values  
in the EA model at zero temperature 
 for almost all $\bm J$ 
 in any dimensions. \\

\noindent
{\bf  Corollary} {\it In the Edwards-Anderson (EA) model, 
RSB in Chatterjee's definition  does not appear  at  zero temperature.}\\

\paragraph{Proof of Corollary.}  Let us evaluate the expectation value of the  site overlap at zero temperature
\begin{eqnarray}
&&\lim_{\beta \to\infty}  \lim_{L\to\infty}  \langle S_{1,2}  \rangle =\lim_{\beta \to\infty}  \lim_{L\to\infty}  
 \frac{1}{|\Lambda_M|} \sum_{i\in \Lambda_M}
 \langle \sigma_i^1 \sigma_i^2\rangle \nonumber \\
 &&=\lim_{\beta \to\infty}  \lim_{L\to\infty}  \frac{1}{|\Lambda_M|} \sum_{i\in \Lambda_M} \langle \sigma_i \rangle^2 
 =  \frac{1}{|\Lambda_M|} \sum_{i\in \Lambda_M} (\sigma^+_i)^2=1, \nonumber  \\
&&\lim_{\beta \to\infty}  \lim_{L\to\infty}  \langle S_{1,2} ^2 \rangle =\lim_{\beta \to\infty}  \lim_{L\to\infty}    \frac{1}{|\Lambda_M|^2} \sum_{i,j\in \Lambda_M}
 \langle \sigma_i^1 \sigma_i^2  \sigma_j^1 \sigma_j^2\rangle \nonumber \\
 &&=\lim_{\beta \to\infty}  \lim_{L\to\infty}  \frac{1}{|\Lambda_M|^2} \sum_{i,j\in \Lambda_M}
 \langle \sigma_i  \sigma_j\rangle^2 \nonumber  \\
 &&= \frac{1}{|\Lambda_M|^2} \sum_{i,j\in \Lambda_M} (\sigma_i^+\sigma_j^+)^2 =1.
 \end{eqnarray}
 These are independent of $M$, and then  
 \begin{eqnarray}
 && \lim_{M \rightarrow \infty }
  \lim_{\beta  \rightarrow \infty } \lim_{L \rightarrow \infty }
\langle S_{1,2}  \rangle  
=  \lim_{M \rightarrow \infty }\lim_{\beta  \rightarrow \infty } \lim_{L \rightarrow \infty }
\langle S_{1,2}  \rangle^2 \nonumber \\
&&= \lim_{M \rightarrow \infty }\lim_{\beta  \rightarrow \infty } \lim_{L \rightarrow \infty }
\langle S_{1,2}^2  \rangle=1. 
\label{end2}
\end{eqnarray}
These and the  dominated convergence theorem imply 
  \begin{eqnarray}
  &&\lim_{M \rightarrow \infty } \lim_{\beta  \rightarrow \infty } \lim_{L \rightarrow \infty }{\mathbb E}
\langle S_{1,2}  \rangle 
 =\lim_{M \rightarrow \infty } \lim_{\beta  \rightarrow \infty } \lim_{L \rightarrow \infty }{\mathbb E}
\langle S_{1,2}  \rangle^2 \nonumber  \\
&&=\lim_{M \rightarrow \infty }\lim_{\beta  \rightarrow \infty } \lim_{L \rightarrow \infty }{\mathbb E}
\langle S_{1,2}^2  \rangle=1. 
\label{end3}
\end{eqnarray}
The variance of the site-overlap vanishes. 
 Also, the expectations of bond-overlap are given by 
\begin{eqnarray}
&& \lim_{M \rightarrow \infty }\lim_{\beta \to \infty} \lim_{L \to \infty}{\mathbb E}\langle R_{1,2} \rangle
= \lim_{M \rightarrow \infty }\lim_{\beta \to \infty} \lim_{L \to \infty}{\mathbb E}\langle R_{1,2} \rangle^2\nonumber\\&&
= \lim_{M \rightarrow \infty }\lim_{\beta \to \infty} \lim_{L \to \infty}{\mathbb E}\langle R_{1,2}^2 \rangle=1.
\end{eqnarray} 
These  complete the proof of Corollary.
$\Box$\\

The following  lemma enable us to prove Theorem. \\

 \noindent
{\bf  Lemma } {\it  Let $f(\sigma)$  be 
an arbitrary uniformly  bounded  real valued function  of spin configuration $\sigma \in \Sigma_L$,  such that $|f(\sigma)| \leq C.$
 For any bond $b \in B_L$ and 
 for almost all $J_{ij}$, the infinite volume limit and the  zero temperature 
 limit of the connected correlation function  vanishes  
\begin{equation}
\lim_{\beta  \rightarrow \infty }\lim_{L \rightarrow \infty }[ \langle \sigma_i \sigma_j  f(\sigma)  \rangle -\langle \sigma_i \sigma_j \rangle \langle f( \sigma )\rangle] =0.
\label{bond}
\end{equation} }

\paragraph{Proof of Lemma }  The derivative of one point function gives 
\begin{equation}
 \frac{1}{\beta}\frac{\partial}{ \partial J_{ij}} \langle f( \sigma) \rangle =  \langle \sigma_i \sigma_j f( \sigma) \rangle -\langle \sigma_i \sigma_j \rangle \langle f(\sigma)\rangle.
\end{equation}
The integration over an arbitrary interval $(J_1,J_2)$ is 
$$
 \frac{1}{\beta}[ \langle f( \sigma) \rangle_{J_2} -  \langle f(\sigma)\rangle_{J_1}]=\int_{J_1} ^{J_2} \hspace{-3mm}dJ_{ij}[  \langle \sigma_i \sigma_j f(\sigma) \rangle -\langle \sigma_i\sigma_j\rangle \langle f(\sigma) \rangle].
$$
Uniform bounds $|f(\sigma)| \leq C$ in the left hand side,
$-2C \leq \langle \sigma_i \sigma_j  f(\sigma) \rangle -\langle \sigma_i \sigma_j \rangle \langle f(\sigma) \rangle \leq 2C
$ on the integrand  in the right hand side,
and the dominated convergence theorem  imply  the following commutativity between the limiting procedure and the integration
\begin{eqnarray}
0&=&\lim_{\beta  \rightarrow \infty } \lim_{L \rightarrow \infty }\int_{J_1} ^{J_2} dJ_{ij}[  \langle \sigma_i \sigma_j  f(\sigma) \rangle -\langle \sigma_i \sigma_j\rangle \langle f(\sigma) \rangle]\\
&=&\int_{J_1} ^{J_2} dJ_{ij} \lim_{\beta  \rightarrow \infty }\lim_{L \rightarrow \infty } [  \langle \sigma_i \sigma_j f(\sigma) \rangle -\langle \sigma_i \sigma_j \rangle \langle f(\sigma) \rangle].
\end{eqnarray}
Since the integration interval $(J_1,J_2)$ is arbitrary, the following limit vanishes
\begin{equation}
\lim_{\beta  \rightarrow \infty }\lim_{L \rightarrow \infty }[  \langle \sigma_i \sigma_j  f(\sigma) \rangle -\langle \sigma_i \sigma_j \rangle \langle f(\sigma) \rangle]=0, 
\end{equation}
for any $\{i,j\} \in B_L$ for almost all $J_{ij} \in {\mathbb R}$.  
$\Box$\\

 Lemma indicates  the following  consistent  property of the spin correlations
 at zero temperature.
Consider  a plaquette ($i,j,k,l)$  with an arbitrary $i\in \Lambda_L$ and
 $j=i+e,k=i+e',l=i+e+e'$ for unit vectors with  $|e|=|e'|=1$.   Lemma for $b=\{i,j\}, \{i,k\}$ and
  $f(\sigma)=\sigma_j\sigma_l,  \sigma_k \sigma_l$ implies
 \begin{eqnarray}
&& \lim_{\beta\to\infty} \lim_{L\to\infty} [\langle \sigma_i  \sigma_j   \sigma_j \sigma_l \rangle - \langle \sigma_i  \sigma_j  \rangle \langle \sigma_j \sigma_ l\rangle]
 =  0,
 \\
&& \lim_{\beta\to\infty} \lim_{L\to\infty} [\langle \sigma_i  \sigma_k   \sigma_k \sigma_l \rangle - \langle \sigma_i  \sigma_k  \rangle \langle \sigma_k\sigma_ l\rangle]  =0.
  \end{eqnarray}
These, $ \sigma_j^2=\sigma_k^2=1$ and eq.(\ref{bond2}) give the consistent property of the spin correlations
 $$ \displaystyle \lim_{\beta\to\infty} \lim_{L\to\infty} \langle \sigma_i  \sigma_j  \rangle \langle \sigma_j\sigma_ l\rangle
 \langle \sigma_l  \sigma_k  \rangle \langle \sigma_k\sigma_ i\rangle =1.$$
  
  \paragraph{Proof of Theorem}  For $M<L,$ note that $\Lambda_M\subset\Lambda_L.$
Eq.(\ref{bond}) in  Lemma  for an arbitrary  bond $\{i,j\} \in B_M$  and $f(\sigma)=\sigma_i \sigma_j$ implies
 \begin{equation}
 \lim_{\beta  \rightarrow \infty }\lim_{L \rightarrow \infty }
  (1-\langle \sigma_i \sigma_j  \rangle^2) = 0.
\label{bond2}
 \end{equation}
Either a ferromagnetic $\langle \sigma_i \sigma_j \rangle =1$ or an antiferromagnetic  $\langle \sigma_i \sigma_j \rangle=-1$ spin correlation appears
  almost surely on any bond $ \{ i,j\}\in B_M$
 for almost all  $\bm J$  at zero temperature.

 For any site $i \in\Lambda_M
 $ and   for $b=\{i,j\}\in B_M$,  
 Lemma and $f(\sigma)=\sigma_i$
 imply
 $$
 \lim_{\beta \to \infty} \lim_{L\to \infty} \langle \sigma_j \rangle =
\lim_{\beta \to \infty} \lim_{L\to \infty}  \langle \sigma_i \sigma_j \rangle \langle \sigma_i \rangle
 $$
 for almost all $\bm J$. For  any sites 
 $i,j\in \Lambda_M$
and $i,j$ are connected by bonds in $B_M$.
Then, the condition  $\sigma_{\bm 0} =1$ given by (\ref{BC}) 
and a spin  correlation $\langle \sigma_i \sigma_j\rangle$ 
fix a spin configuration  $\sigma^+\in \Sigma_M$  uniquely at zero temperature for any $M$.
This spin configuration $\sigma^+$
gives 
$$
\lim_{\beta \to \infty} \lim_{L\to \infty} \langle f(\sigma) \rangle  = f( \sigma^+),
$$
for a real valued function $f(\sigma)$  of $\sigma \in \Sigma_M$.
This completes the proof. $\Box$

Note that the  ferromagnetic order parameter $m$ and the  spin glass order parameter $q$ are
$m=0, q=1$  in   
 the spin glass phase and  $m\neq 0, q=1$ in   
 the ferromagnetic phase at zero temperature.

\paragraph{Discussions}
In the present paper, it has been proven that the zero temperature  infinite volume Gibbs state gives a unique   spin configuration
in  the Edwards-Anderson model with  continuous random exchange interactions
 in any dimensions.
In this state, the  site- and bond-overlap are concentrated at their 
maximal values. 

Here, we  comment on the RSB in mean field spin glass  models. 
For example,
the Hamiltonian of the Sherrington-Kirkpatrick model 
defined by 
\begin{equation}
H_N(\sigma
):=-\sum_{1\leq i<j\leq N}  \Big(\frac{J_{i,j} }{\sqrt{N}}+\frac{J_0}{N} \Big)\sigma_i\sigma_j,
\end{equation}
is a function of the spin configuration of $N$ spins.
In  this model,  Lemma  for an arbitrary $1\leq i,j\leq N$  and a bounded real valued function  $f(\sigma)$ gives 
\begin{equation}
 \lim_{\beta \to \infty}
  \lim_{N \to \infty}\frac{1}{\sqrt{N}} [\langle \sigma_i \sigma_j  f(\sigma) \rangle -
 \langle \sigma_i \sigma_j \rangle \langle   f(\sigma) \rangle ] =0.
\end{equation}
The identity (\ref{bond}) in Lemma  cannot be obtained from the above relation in the infinite-volume limit $N\to\infty$. 
Therefore, the present argument does not rule out
 the RSB  in mean field spin glass models. 
 The RSB property of the  Parisi measure  for the SK model  is proven by Auffinger, Chen and Zeng. 
 For any $\epsilon, \eta >0$ and any $u \in (0,1)$, there exist two spin configurations $\sigma^1, \sigma^2$  with $\displaystyle  S_{1,2}=\sum_{i=1}^N \sigma_i^1\sigma_i^2$ and $K >0$,  such that
 \begin{eqnarray}
&& P[S_{1,2} \in (u-\epsilon, u+\epsilon),  H_N(\sigma^1), H_N(\sigma^2) \leq \min_\sigma H_N(\sigma)+N \eta  ] \nonumber \\
 &&\geq 1- K e^{-\frac{N}{K}}. \label{SK}
 \end{eqnarray}
 Here, we discuss the energy gap above the unique ground state $\sigma^+ \in \Sigma_L^+
 $ for an arbitrary fixed $\bm J$
  in the EA model, where $\Sigma_L^+ ( \subset \Sigma_L)$ denotes a set of  spin configurations $\sigma$ satisfying  the condition $\sigma_{\bm 0}=1$.
For an arbitrary subset $ S (\subset \Lambda_L \setminus \{{\bm 0}\})$, define $\tau^S \in \Sigma_L^+$ by   ${\tau^S}_i = -\sigma^+_i$ for  $i \in S$ and 
$\tau_i = \sigma^+_i$ for 
$i \in S^c$.  
The  boundary $\partial S$ of $S$ is a set of bonds defined by  
 $$\partial S:=\{ \{i,j\} \in B_L| i \in S, j \in {S}^c  \ {\rm or}  \  j \in S, i \in {S}^c\}.$$ 
 The energy gap of  the spin configuration $\tau^S$
 $$H_L(\tau^S,{\bm J})-H_L(\sigma^+, {\bm J})=2\sum_{b \in \partial S} J_{ij} \sigma_i^+ \sigma_j^+,$$
 is always positive for any $S \subset \Lambda_L \setminus \{{\bm 0}\}$.
 Define a condition on $\bm J$ for positive energy gaps
 $$
 C({\bm J}) := \bigcap_{S\subset \Lambda_L \setminus \{\bm 0\}}
\{ {\bm J} | \sum_{b \in  \partial S
} J_{ij} \sigma_i ^+\sigma_j^+ > 0 \},
 $$
 Define an indicator $I$ by  $I[{\rm true}]=1$ and $I[{\rm false}] =0$, and define a
  conditional  probability of any event e
 under the condition $C({\bm J})$
\begin{eqnarray}
&&
P[{\rm e}\ |\ C(\bm J)]
=
\frac{{\mathbb E}
I [ {\rm e} ]  \ I[C(\bm J)]}{{\mathbb E}  I[\ C(\bm J) \ ]}. \nonumber 
\label{condprob}
\end{eqnarray}
 This conditional probability 
predicts  that the energy gap of the spin configuration  $\tau^S$ 
 is proportional to $|\partial S|$ as well as the pure
Ising model. 
The overlap between $\sigma^1=\sigma^+$  and  $\sigma^2=\tau^S$ is given by
$$
S_{1,2} = \frac{1}{|\Lambda_M|} \sum_{i\in \Lambda_M} \tau ^S_i \sigma^+ _i = 1- \frac{2|S|}{|\Lambda_M|}.
$$
It is proven  that the appearance of a spin configuration with $S_{1,2} < 1$ and  an infinitesimal energy gap is a rare event \cite{I2}.
This property  differs from that in the SK model given by (\ref{SK}), and
the fluctuation from $S_{1,2}=1$ near zero temperature
with  $|S| =r |\Lambda_M|$ for  $r>0$ is suppressed by the property of  energy gap above the unique ground state
in the EA model. 
Therefore, the droplet picture by Fisher and Huse \cite{FH} is correct picture to understand the EA model in sufficiently low temperature.  
\\

Acknowledgment     

It is pleasure to thank  R. M. Woloshyn for careful reading of the manuscript.
I would like to thank T. Koma for helpful discussion in the early stage of this work.
I appreciate  informations of the SK model  M. Moore and Q. Zeng have given me.
I am grateful to  S. Suzuki,   K. Sato,   K. Horie,  H. Shimajiri  and  Y. Sakamoto for helpful discussions.


\vspace{2cm}

\noindent
{\bf {\Large  Errata :Uniqueness of ground state in the  Edwards-Anderson \\ \hspace{2cm} spin glass model
\small{[J. Phys. Soc. Jpn. {\bf 90}, 033002(2021)]}}}\\

 Theorem in the present paper$^{1)}$ is valid for the Edwards-Anderson (EA) model 
in a  finite lattice. 
Therefore,  Eq.(8) should be corrected  to the following  
$$
\lim_{\beta \to \infty}  \langle f(\sigma )\rangle= f( \sigma^+),  \ \  \ \ \ \ \ \ \ \ \ \ \ (8)
$$ 
to provide a rigorous result, since
the convergence of $\langle f(\sigma) \rangle$ in 
the infinite volume limit is assumed in the proof of Theorem.
If symbols of the infinite volume limit 
are removed from all equations throughout this paper, they become correct because of 
 the validity of Lemma for each system size $L$ in the zero temperature limit.  
 This implies that  the reverse order of two limiting operations  gives 
  correct identities 
$\displaystyle \lim_{L\to\infty}\lim_{\beta\to\infty} \langle  \sigma_i\sigma_j  f(\sigma)\rangle 
- \langle  \sigma_i\sigma_j \rangle\langle f(\sigma)\rangle=0
$ in Lemma  and 
$\displaystyle \lim_{L\to\infty}\lim_{\beta \to \infty} \langle S_{1,2}\rangle
=1=\lim_{L\to\infty}\lim_{\beta \to \infty} \langle R_{1,2}\rangle$ in Corollary.
 Even  after taking   the zero temperature limit,  however, it is still difficult to evaluate  the  system size dependence  of
 $\langle f( \sigma )\rangle$. 
Then, its  infinite volume limit is not proven rigorously in the present paper.

In Discussions, the statement 
``It is proven that the appearance of a spin configuration with $S_{1,2} < 1$ and an infinitesimal energy gap is a rare event$^{28)}$ "
 claims  as if  absence of replica symmetry breaking (RSB) were proven in the EA model in  low temperatures. 
Although the significantly small variance of the energy gap  
and the law of large numbers  verify  this statement in the EA model in a finite lattice, 
 this cannot immediately imply that 
 the variance of $S_{1,2}$ vanishes in the infinite volume limit. 
To conclude the absence of RSB
in  the EA model in low temperatures, it is necessarry to prove that the variance of  $S_{1,2}$
  vanishes in the infinite volume limit.
It should be studied  still whether RSB occurs in the EA model or not.  
\\

{\bf Acknowledgment}\\
  It is pleasure to thank  H. Nishimori,  H. Tasaki and H. Yoshino for helpful discussions.\\

{\bf  References}\\1) \  C. Itoi, J. Phys. Soc. Jpn. {\bf 90}, 033002(2021)

\begin{thebibliography}{99}
%

\bibitem{Pr} G. Parisi, J. Phys. A {\bf13}  L-115 (1980).

\bibitem{SK} D. Sherrington and S. Kirkpatrick, Phys. Rev. Lett. {\bf 35}  1792
(1975).

\bibitem{T} M. Talagrand,   Ann. of Math. {\bf 163} 221
(2006)

\bibitem{T2} M. Talagrand,  ``Mean field models for spin glasses" Springer, Berlin (2011).

\bibitem{ACZ} A. Auffinger, W-K. Chen and Q. Zeng, arXiv.1703.06872m, (2017).

 \bibitem{EA} S. F.  Edwards and P. W. Anderson, J. Phys. F: Metal Phys. {\bf 5} 965
(1975).
\bibitem{P2} G. Parisi, Phys. Rev. Lett. {\bf 50}, 1946 (1983).

\bibitem{FH}  D. S. Fisher and D. A. Huse,   Phys. Rev. Lett. {\bf 56} 1601 (1986).

\bibitem{MPRRZ} E. Marinari,  G. Parisi, F. Ricci-Tersenghi,  J. J. Ruiz-Lorenzo, and F.  Zuliani, 
J.  Stat. Phys.  {\bf  98}, 5973, (2000).

\bibitem{KPY} H. G. Katzgraber, M. Palassini, and A. P. Young, Phys. Rev. B {\bf 63}, 184422 (2001).

\bibitem{BCFGM} R. A. Ba\~nos, A. Cruz, L. A. Fernandez, J. M. Gil-Narvion, A. Gordillo-Guerrero, M. 
Guidetti, A. Maiorano, F. Mantovani, E Marinari and   V Martin-Mayor, J. Stat. Mech. {\bf 2010} 06026 (2010).

\bibitem{WWL}  W.Wang ,  M.  Wallin and J.  Lidmar, Phys. Rev. Research  {\bf 2}, 043241 (2020).

\bibitem{YK} A. P. Young and H. G. Katzgraber, Phys. Rev. Lett. {\bf 93}, 207203 (2004).

\bibitem{YKM} B. Yucesoy, H. G. Katzgraber and J. Machta,  Phys. Rev. Lett. {\bf 109}, 177204 (2012).





\bibitem{NS1} H. Nishimori and D. Sherrington, AIP Conference Proceedings 553, 67 (2001).

\bibitem{N} H. Nishimori, ``Statistical Physics of Spin Glasses and Information Processing: An Introduction"
Oxford university press (2001)

\bibitem{AD} L-P. Arguin and M. Damron, Ann. Inst. H. Poinca\'re Probab. Statist.  {\bf 50} 28 (2014). 

\bibitem{C2} S. Chatterjee, Commun. Math .Phys. {\bf 337}93
(2015)

 \bibitem{FKG} C. M. Fortuin, P. W. Kasteleyn and J. Ginibre,  Commun. Math. Phys. {\bf 22} 89
 (1971). 

\bibitem{GG} S. Ghirlanda and F. Guerra  J. Phys. A{\bf 31} 9149
(1998).

 \bibitem{AC} M. Aizenman and P. Contucci, J. Stat. Phys. {\bf 92}765
 (1998).




\bibitem{K} F. Krzakala, F. Ricci-Tersenghi and L. Zdeborova, Phys. Rev. Lett. {\bf 104} 207208
 (2010) 


\bibitem{I} C. Itoi and Y. Utsunomiya,  J. Math. Phys. {\bf 61} 073302 (2020).


\bibitem{G2} F. Guerra, J. Phys: Conf. Series {\bf 442}012013 (2013).






	 


\bibitem{G} F. Guerra,   Commun. Math. Phys. {\bf 233} 1, (2003).





 \bibitem{Pn} D. Panchenko, Compt. Read. Math. {\bf 348} 189
 (2010).

 
\bibitem{NS} C. M. Newman and D. L. Stein, Phys. Rev. Lett. {\bf 76} 515(1996)
.




 
 

\bibitem{I2} C. Itoi, in preperation.

\end{thebibliography}
\end{document}